\documentclass[preprint, review, 3p, 12pt]{elsarticle}
\usepackage[american]{babel}
\usepackage[T1]{fontenc}
\usepackage{amsmath}
\usepackage{amssymb}
\usepackage{mathtools}
\usepackage{siunitx}
\usepackage{miller}
\usepackage{booktabs}
\usepackage{makecell}
\usepackage{xcolor}
\usepackage{enumitem}
\usepackage[list=true, font=large, labelfont=bf, labelformat=brace, position=top]{subcaption}
\usepackage{soul}
\usepackage{url}
\usepackage{varioref}
\usepackage{hyperref}
\usepackage[nameinlink]{cleveref}
\definecolor{tud0d}{RGB}{83,83,83}
\definecolor{tud0c}{RGB}{137,137,137}
\definecolor{tud0b}{RGB}{181,181,181}
\definecolor{tud0a}{RGB}{220,220,220}

\definecolor{tud1a}{RGB}{93,133,195}
\definecolor{tud2a}{RGB}{0,156,218}
\definecolor{tud3a}{RGB}{80,182,149}
\definecolor{tud4a}{RGB}{175,204,80}
\definecolor{tud5a}{RGB}{221,223,72}
\definecolor{tud6a}{RGB}{255,224,92}
\definecolor{tud7a}{RGB}{248,186,60}
\definecolor{tud8a}{RGB}{238,122,52}
\definecolor{tud9a}{RGB}{233,80,62}
\definecolor{tud10a}{RGB}{201,48,142}
\definecolor{tud11a}{RGB}{128,69,151}

\definecolor{tud1b}{RGB}{0,90,169}
\definecolor{tud2b}{RGB}{0,131,204}
\definecolor{tud3b}{RGB}{0,157,129}
\definecolor{tud4b}{RGB}{153,192,0}
\definecolor{tud5b}{RGB}{201,212,0}
\definecolor{tud6b}{RGB}{253,202,0}
\definecolor{tud7b}{RGB}{245,163,0}
\definecolor{tud8b}{RGB}{236,101,0}
\definecolor{tud9b}{RGB}{230,0,26}
\definecolor{tud10b}{RGB}{166,0,132}
\definecolor{tud11b}{RGB}{114,16,133}

\definecolor{tud1c}{RGB}{0,78,138}
\definecolor{tud2c}{RGB}{0,104,157}
\definecolor{tud3c}{RGB}{0,136,119}
\definecolor{tud4c}{RGB}{127,171,22}
\definecolor{tud5c}{RGB}{177,189,0}
\definecolor{tud6c}{RGB}{215,172,0}
\definecolor{tud7c}{RGB}{210,135,0}
\definecolor{tud8c}{RGB}{204,76,3}
\definecolor{tud9c}{RGB}{185,15,34}
\definecolor{tud10c}{RGB}{149,17,105}
\definecolor{tud11c}{RGB}{97,28,115}

\definecolor{tud1d}{RGB}{36,53,114}
\definecolor{tud2d}{RGB}{0,78,115}
\definecolor{tud3d}{RGB}{0,113,94}
\definecolor{tud4d}{RGB}{106,139,55}
\definecolor{tud5d}{RGB}{153,166,4}
\definecolor{tud6d}{RGB}{174,142,0}
\definecolor{tud7d}{RGB}{190,111,0}
\definecolor{tud8d}{RGB}{169,73,19}
\definecolor{tud9d}{RGB}{156,28,38}
\definecolor{tud10d}{RGB}{115,32,84}
\definecolor{tud11d}{RGB}{76,34,106}

\usepackage{tikz}
\usepackage{pgfplots}
\usepackage{pgfplotstable}
\pgfplotsset{compat=newest}
\pgfplotsset{
        layers/my layer set/.define layer set={
            background,
            main,
            foreground
        }{
        },
        set layers=my layer set,
    }

\usepgfplotslibrary{external}
\tikzsetexternalprefix{tikz/}
\usepgfplotslibrary{groupplots}
\usepgfplotslibrary{fillbetween}
\usetikzlibrary{automata}
\usetikzlibrary{positioning}
\usetikzlibrary{calc}
\usetikzlibrary{plotmarks}
\usetikzlibrary{arrows}
\tikzset{external/system call={
    lualatex
    \tikzexternalcheckshellescape -halt-on-error -interaction=batchmode 
        -jobname "\image" "\texsource"
        }}
\pgfkeys{/pgf/number format/.cd,
    1000 sep={\,},
    min exponent for 1000 sep=4}
\pgfplotsset{unit code/.code 2 args={\si{#1#2}}}
\pgfplotsset{
    legend image with text/.style={
        legend image code/.code={%
            \node[anchor=center] at (0.3cm,0cm) {#1};
        }
    },
}

\biboptions{sort&compress}
\hypersetup{
pdfborder={0 0 0},
breaklinks=true,}

\begin{document}

\begin{frontmatter}

    \title{Mechanical Characterization of Superelastic NiTi Nanofoams by Molecular Dynamics Simulations}

    \author[1]{Arne J. Klomp 
    \fnref{fn1}}
    \ead{klomp@mm.tu-darmstadt.de}
    \author[1]{Karsten Albe}

    \address[1]{Materials Modelling Division, TU Darmstadt, Otto-Berndt-Str. 3, 64287 Darmstadt, Germany}

    \begin{abstract}
        Nanoporous metals or nanofoams are a promising material class that is considered for sensing, actuation, and catalysis.
        To date, they mostly based on simple noble metals such as nanoporous gold, which exhibit peculiar stress-strain response different from the bulk material.
        At the same time bulk alloys such as NiTi feature a reversible martensitic phase transition giving rise to interesting shape memory and superelastic effects.
        Combining the rich mechanics of NiTi with the geometrical features of a nanofoam is expected to improve the mechanical performance of this material.
        In this atomistic study we explore the behavior of a NiTi nanofoam at varying temperature and its reaction to (cyclic) compression.
        Using molecular dynamics simulations we track the microscopic processes enabling reversible deformation as well as the mechanical failure mechanisms of the NiTi nanofoam.
    \end{abstract}
    
    
\end{frontmatter}


\section{Motivation}

    Metallic nanofoams of noble metals are a well investigated class of materials that have been enabled by modern synthesis techniques.
    They possess a unique combination of functional and mechanical properties that make them valuable for sensor, catalyst, and actuator applications \cite{Weissmuller2003}.
    Recently, it has been shown that even less noble metals and alloys can be transformed into nanofoams \cite{Okulov2018}.
    Yet, the materials studied to date are all simple ductile metals.
    By introducing the rich mechanical properties of superelastic/shape-memory alloys into this configuration, we expect new flavors of deformation behavior.
    Ultimately, nanometer-sized shape memory foams can open a new window for applications of superelastic functional materials focussing on actuation, damping, and elastocalorics.
    
    To this end, we take the most well investigated superelastic alloy NiTi and produce foams with nanometer-sized features from this material.
    Our approach combines an existing bulk material (NiTi) with a well studied geometrical arrangement (nanofoam) in a computational simulation.
    Using a model that represents material configuration as well as thermomechanical behavior on an atomic scale, the response to thermal and mechanical load are evaluated \cite{Ko2015}.
    We aim to predict macroscopic material behavior and ultimately stimulate experimental work.

\section{Introduction}
\label{sec:introduction}

    As a basis for this work we review literature approaching the problem from two sides:
    Metallic nanofoams and macroscopic NiTi foams.

    \subsection{Metallic nanofoams}
            
        Modern technological advancements approaching increasingly small device dimensions have revealed the potential of nanostructured materials for a variety of applications.
        In metallic nanofoams the structural features of a foam have been brought to the nanoscale.
        A significant amount of research has been performed to reveal the distinct material behavior of gold nanofoams (\textit{nanoporous gold} or \textit{NPG}).

        These intriguing mechanical properties have been recently reviewed in Refs. \cite{Jin2018, Ngo2015} and include:
        \begin{enumerate}[label=\alph*)]
            \item very low initial stiffness with pronounced stiffening during compression;
            \item immediate onset of yielding upon loading;
            \item homogeneous deformation to very large strains in compression;
            \item high strain-hardening coefficient;
            \item strong formation of dislocations and stacking faults under compression; and
            \item Gibson-Ashby scaling laws for Young's modulus and yield strength strongly overestimate these values.
        \end{enumerate}

        To this point, research has focused mostly of foams of simple noble metals.
        A characteristic feature of NPG is that every little change in stress applied during initial loading leads to plastic compliance in the material \cite{Ngo2015}.
        Since deformation can be accommodated by the martensitic phase transformation in NiTi, we ask how the possible superelastic/shape-memory effect of the base alloy affects the mechanical properties of a NiTi nanofoam?
        Does a NiTi nanofoam behave more like nanoporous noble metals or does it behave more like a bulk superelastic NiTi alloy?
        These questions are especially important in the light of actuation and damping applications.

    \subsection{Porous NiTi}
    
\begin{figure*}
            \centering
            \includegraphics{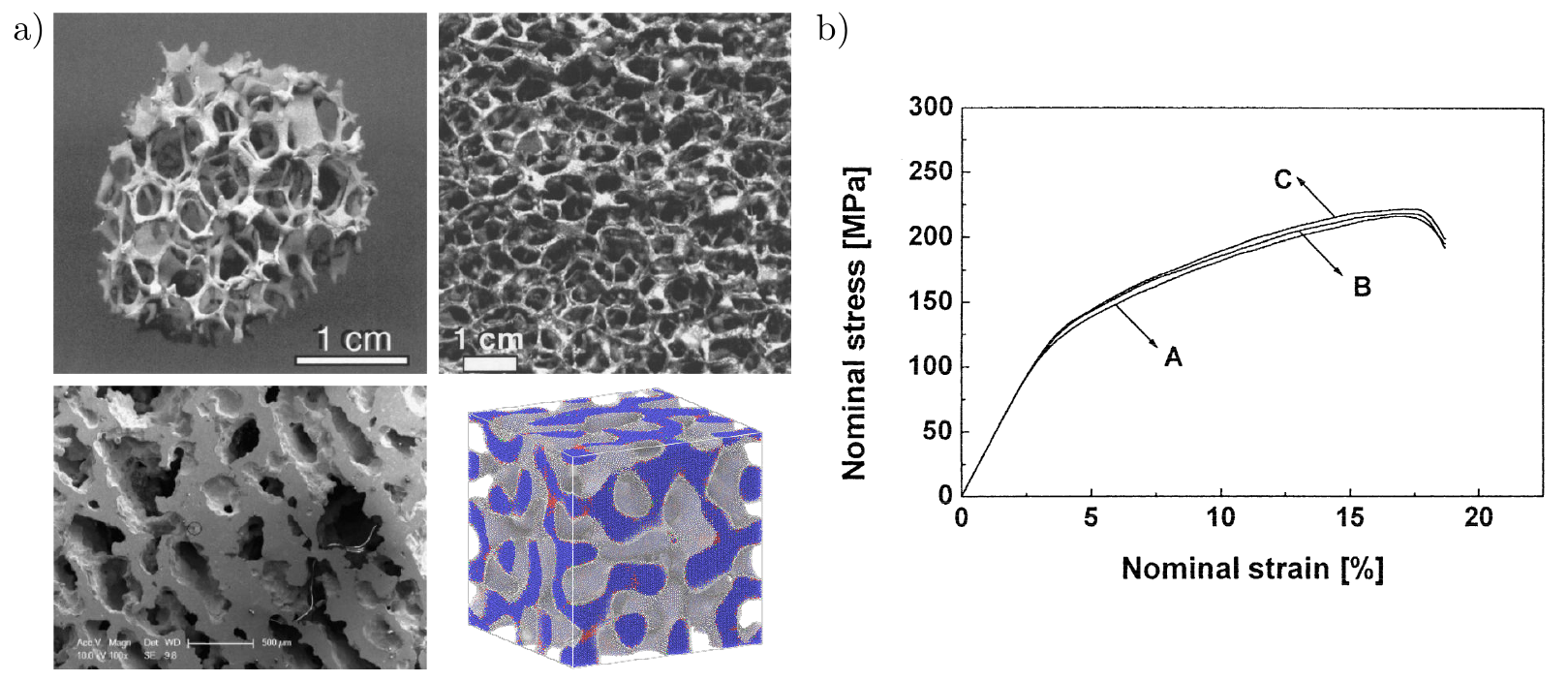}
            \caption{
            (a)
            Foams from experiment and simulation.
            Top row: NiTi foams produced by Grummon \emph{et al.} with structural features in the millimeter range \cite{Grummon2003}.
            Reproduced with permission from Appl. Phys. Lett. 82, 16 (2003). Copyright 2003, AIP Publishing.
            Bottom left: NiTi foam produced by Guo \emph{et al.} with structural features in the micrometer range \cite{Guo2009}.
            Reproduced with permission from Mater. Sci. Eng. A 515, 1-2 (2009). Copyright 2009, Elsevier B.V..
            Bottom right: Simulated NiTi foam from this work with structural elements in the nanometer range.
            (b)
            Typical stress-strain curve of a NiTi foam with \SI{54.2}{\percent} porosity \cite{Li2002}.
            Stress and strain are given for compression with rates of \SI{1e-3}{\per\second} (A), \SI{1e-2}{\per\second} (B), \SI{1e-1}{\per\second} (C).
            Reproduced with permission from J. Alloys Compd 345, 1-2 (2002). Copyright 2002, Elsevier B.V..                
            }
        \label{fig:foam_experimental}
        \end{figure*}

        Nanosized foam-like structures of NiTi are new to literature.
        There is, however, a considerable number of publications on macroscopic porous NiTi structures, which are discussed as promising candidates for implants \cite{Bansiddhi2008}.
        A variety of powder metallurgical syntheses have been established \cite{Yuan2006, Panigrahi2006, Sadrnezhaad2006, Zhu2005, Bertheville2006, Xu2015, Zhao2005, Nemat-Nasser2005, Zhang2015, Entchev2004, Yuan2004, Yuan2005, Greiner2005, Wu2006, Zhang2008, Sugiyama2007, Li2000, Li2002, Yeh2004, Lagoudas2002, Kim2004, Chu2005, Li2006, Guo2009, Bassani2009, Grummon2003, Bansiddhi2007, Bram2011, Aydogmus2012, Hosseini2014, Sakurai2006}, and investigations have been supplemented by micromechanical modeling \cite{Entchev2004,Nemat-Nasser2005}.
        They cover a wide range of porosities from \SI{6}{\percent} to \SI{96}{\percent}, open porosity fractions of up to \SI{95}{\percent} and pores sizes from few tens of micrometers to millimeters, see \Cref{fig:foam_experimental} (a). 

        In compression tests, three stages in the stress-strain curve can be distinguished \cite{Li2002, Hosseini2014, Xu2015}.
        At low strains porous NiTi behaves according to linear elasticity which is interpreted as simple elastic compression or bending of foam cell walls.
        As strain increases a yielding phase sets in where the slope in the stress-strain curve is decreased due to the collapse of foams cells.
        The yielding phase ends when densification has approached a critical limit and fracture occurs.
        If pore geometry is not regular and has sharp edges, stress concentrations enhance localized phase transformation and promote fracture \cite{Lagoudas2002,Yuan2005,Zhu2005}.
        Also, the smaller the pores, the stronger and stiffer the material and the more abrupt the temperature driven phase transformations \cite{Kim2004, Lagoudas2002}.
        Discussion, whether the Gibson-Ashby scaling laws for elastic modulus and strength are obeyed is still ongoing \cite{Gibson1982, Greiner2005,Hosseini2014, Bansiddhi2007}.

        Upon removal of compressive loading it was generally found that, even above the austenite finish temperature $A_f$, only low strains around \SI{2}{\percent} can be recovered completely (compared to up to \SI{8}{\percent} for bulk NiTi) \cite{Lagoudas2002, Zhu2005, Zhang2008, Bram2011, Zhang2015}.
        The larger the porosity, the worse the strain recovery \cite{Bansiddhi2007, Xu2015, Zhang2015}.
        At larger strains the recovery is decreased but can be enhanced at increased temperatures or with annealing after unloading \cite{Sakurai2006, Aydogmus2012}.
        However, a few studies even find that porous NiTi structures are superior to bulk NiTi in terms of total recovered strain upon unloading \cite{Greiner2005, Sakurai2006}.

        In most of the above experimental studies an appropriate heat treatment or selection of the temperature range fostering superelastic behavior is missing and phase fractions during mechanical loading are not measured at all.
        Thus, the phase transformation and failure mechanisms remain largely unclear.
        In addition, many synthesized samples suffer from irregular pore geometry due to difficult sintering.
        Consequently, the results in literature do not draw a clear picture of the mechanical behavior of porous NiTi.
        With MD simulations these drawbacks are overcome and analysis of the microstructure and phase composition is significantly simplified.

    \subsection{Nanostructured NiTi}

        Since the shape-memory/superelastic effect is intrinsic to the material, down-scaling to the nanometer range is a promising approach.
        It is generally accepted that decreasing feature size (where \textit{feature} can be a nanopillar, a grain in a polycrystal, a wire or a freestanding particle) alters the thermomechanical properties of NiTi \cite{Frick2007, Waitz2009, Mutter2011, Mirzaeifar2014, Ko2017, Chen2018a, Wang2018, Zhang2018}.
        As the local length scale drops below \SI{50}{\nano\meter} to \SI{2.4}{\nano\meter}, the austenite phase is stabilized by a decrease in austenite finish $A_f$ and martensite start $M_s$ temperatures.
        Additionally, superelastic and shape-memory effect decrease along with the hysteresis, i.e., the hysteresis loop becomes slimmer.

        As noted by Ko \emph{et al.} the change in magnitude of transformation temperature, hysteresis and superelastic effect are strongly connected to the stabilization of austenite because of mechanical constraints at surfaces and interfaces \cite{Ko2017}.
        Additionally, it has been found that the deterioration in strain recovery during the first loading cycle is due to residual martensite, disordered phase accumulation and plastic deformation localized at interfaces \cite{Wang2018}.

        We will, thus, investigate if there is indeed a stabilization of the austenite phase and if the transformation temperatures change.
        Possible property improvements or degradations at the nanoscale will be highlighted.

\section{Methods}

    \subsection{Creation of nanofoams}
                
        Classically, virtual nanofoams for simulation have been created by spinodal decomposition and using the resulting structure as a template for the sample \cite{Ngo2015}.
        A more efficient and equally suited method to create open porous structures very similar to those resulting from spinodal decomposition is the wave-modulation method \cite{Soyarslan2018}, which we use here.
        As a simple measure for the relevant length scale in a nanofoam we use the average ligament diameter $L$.
        It is estimated using the method from Ref. \cite{Huber2014, Ngo2015};
        \begin{align}
            L =  1.63 * \frac{(1.25 - \phi)*[1.89 + \phi*(0.505 + \phi)]}{A_{\text{surface}}/V_{\text{solid}}} \quad ,
        \end{align}
        which relies on the solid volume $V_{\text{solid}}$, the total volume of the simulation cell $V_{\text{cell}}$, their fraction $\phi = V_{\text{solid}} / V_{\text{cell}}$ and the surface area $A_{\text{surface}}$.

        To investigate the mechanical stability of nano-sized foam structures we create nanofoams of different solid phase fraction $\phi = 0.2, 0.3, 0.4, 0.5 \text{,~and~} 0.6$ as perfect B2 single crystalline nanofoams.
        The crystallographic primary axes coincide with the simulation cell axes.
        The average ligament diameter $L$ is \SI[separate-uncertainty]{4.02 +- 0.04}{\nano\meter} for all the samples, and the number of atoms ranges from \num{800000} to \num{1260000} for increasing $\phi$.
        Equilibration of the virtual samples is performed for at least \SI{100}{\pico\second} at \SI{460}{\kelvin} ensuring austenitic equilibrium structures.
    
    \subsection{Interatomic potential \& software}
        Classical molecular dynamics (MD) simulations using a second-nearest neighbor modified embedded atom model (2NN MEAM) potential are employed \cite{Ko2015}.
        This interatomic potential shows good approximation of mechanical, thermodynamic and structural properties of NiTi.
        Most importantly it has a good representation of the stress-driven structural transformation from austenite (B2 structure) to martensite (B19' structure) phase and vice versa.
        Compared to micromechanical modeling, atomistic modeling makes localized atomic-level effects traceable.

        As simulation software the Large-scale Atomic/Molecular Massively Parallel Simulator has been used \cite{Plimpton1995,Thompson2022}.
        Structure identification and related analyses are performed with the software Ovito \cite{Stukowski2010,Larsen2016}.
        For Ovito's polyhedral template matching a cut-off of $\text{RMSD}_{\text{max}} = 0.35$ is used.

\section{Results}

    In the following the thermomechanical behavior of NiTi nanofoams is explored systematically.
    First, the equilibrium structure of the pristine foam above its austenite finish temperature is described and benchmarked against bulk NiTi and nanowires.
    Second, temperature cycles spanning from almost \SI{0}{\kelvin} to well above the austenite finish temperature $A_f$ are conducted for bulk, nanowire and foam.
    Third, the nanofoam is subjected to compressive loading and unloading.
    Last, cyclic loading is applied to the foam and the hysteretic and fatigue behavior is tracked.

    \subsection{Equilibrium above \(A_f\)}

        In order to correctly pinpoint the appropriate temperatures for simulating the NiTi nanofoam, especially $A_f$, the transformation temperatures of bulk NiTi and nanowires are evaluated.
        Moreover, it is known from NPG, that even in the absence of applied outer mechanical loading, small portions of the foam are stressed by the surface tension to a degree, that plastic deformation occurs \cite{Ngo2015}.
        Because NiTi can react to mechanical load by phase transformation, we will explicitly monitor the phase composition in the absence of external load.

\begin{figure}[htb]  
            \centering
            \includegraphics{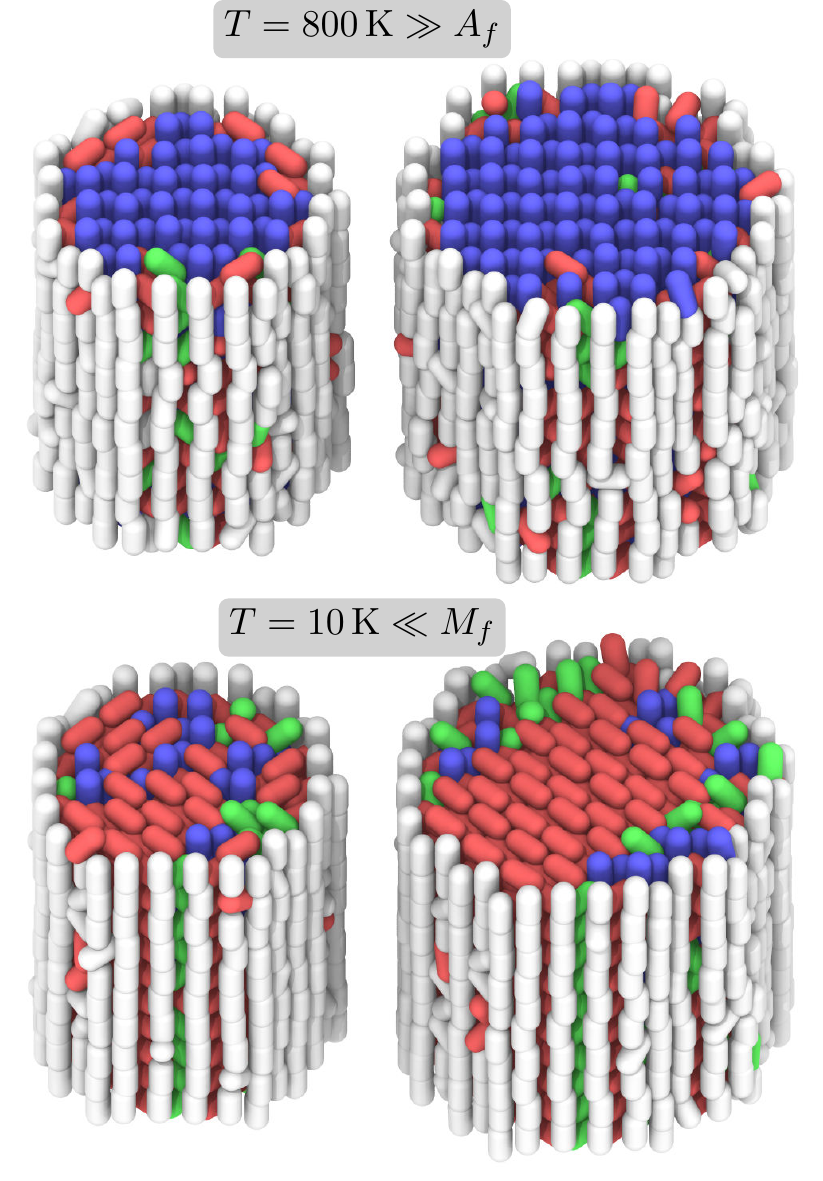}
            \caption{
                Arrangement of phases in small nanowire at different temperatures.
                Nanowires with small diameters (\SI{2.37}{\nano\meter} (left) and \SI{2.98}{\nano\meter} (right)) have been initialized in B2 structure and equilibrated (top).
                After cooling to \SI{10}{\kelvin}, martensite has formed.
                Austenite is represented in blue and martensite in red and orientation of the ellipsoids indicates the orientation of the crystallographic axes.
                }
        \label{fig:domain_boundaries_small_wire_b2_T_cycle}
        \end{figure}

        \paragraph{\textbf{Bulk and nanowire}}
        For the bulk baseline a fully periodic block of NiTi samples with approx. \num{54000} atoms was subjected to a temperature cycle.
        Additionally, round nanowires with varying diameters cut from a perfect B2 crystal
        are subjected to the same temperature cycle.
        Representative examples of nanowires above $A_f$ (top) and below $M_f$ (bottom) are shown in \Cref{fig:domain_boundaries_small_wire_b2_T_cycle}.

        \begin{figure}[htb]  
            \centering
            \includegraphics{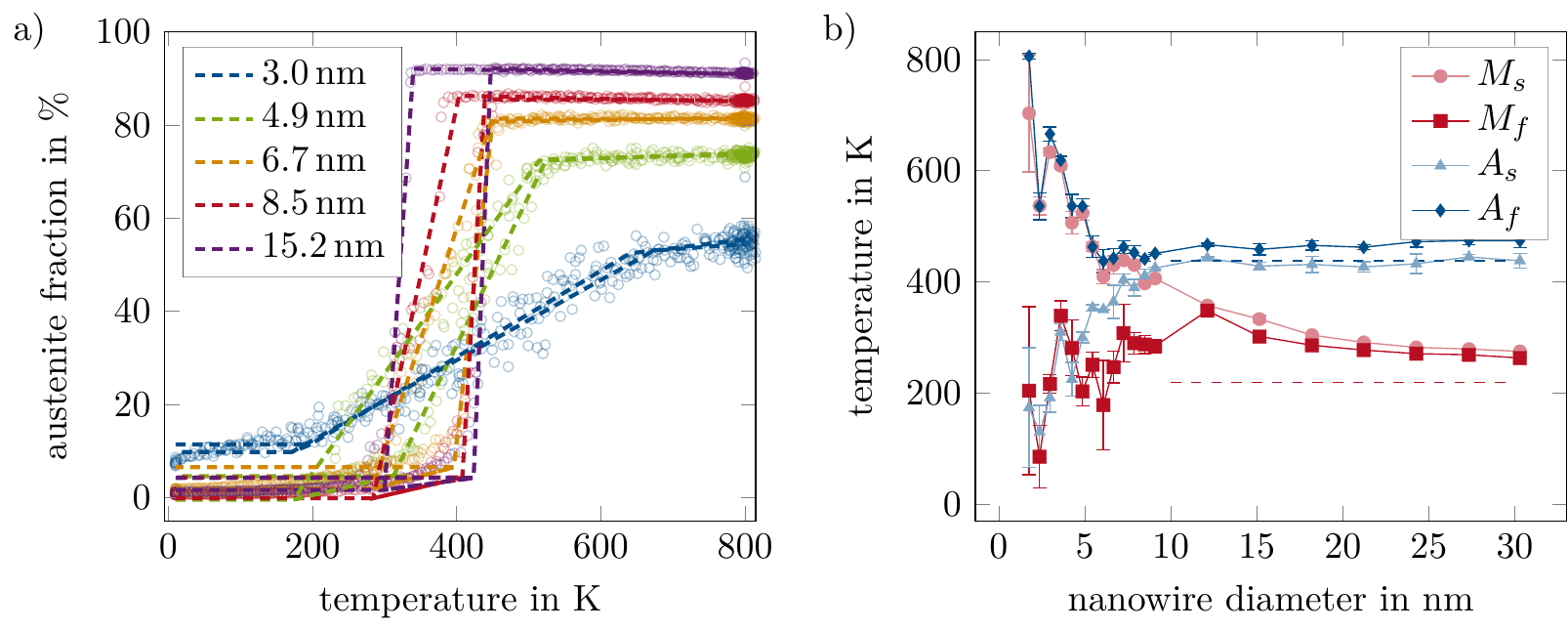}
            \caption{
                (a) Change of austenite of phase fraction in nanowires of varying size.
                The dashed lines indicate piecewise-linear fits that are used to extract the characteristic temperatures.
                (b) Size-effect of nanowire transformation temperatures with B2 starting structure.
                Austenite finish $A_f$ and martensite finish temperatures $M_f$ for round nanowires with varying radius. Dashed lines show the transformation temperatures for bulk NiTi.
                }
        \label{fig:nanowire_phases_temp_cycle}
        \label{fig:nanowire_size_effect}
        \end{figure}

        Examples for the change of austenite phase fraction during a heating and cooling cycle are shown in \Cref{fig:nanowire_size_effect} (a).
        To extract the phase transformation temperatures we use a piecewise-linear fitting algorithm as indicated by the dashed lines.
        The results for austenite start $A_s$, austenite finish $A_f$, martensite start $M_s$ and martensite finish $M_f$ temperatures are shown in \Cref{fig:nanowire_size_effect} (b).

        The martensite start temperature  $M_s$ rises and the austenite start temperature $A_s$ falls as nanowire size decreases while the finish temperatures $M_f$ and $A_f$ behave vice versa.
        The temperature window in which the phase transformation occurs is, thus, significantly widened and the transition becomes \textit{smeared}.
        For large nanowire diameters the bulk values for the transformation temperatures are approached.
        At the same time the spread between start and finish temperatures for the phase transformations decreases indicating more abrupt phase transitions.
        Put in other words, as size decreases and surface effects increase in importance, more martensite phase is retained at high temperatures and more austenite phase remains untransformed at low temperatures.

        Visual inspection of the simulated samples reveals that the martensite formation upon cooling starts inside the nanowire almost homogeneously.
        On the contrary, upon heating the austenite nucleates at the surfaces and the phase transformation progresses into the wire \cite{Ko2017}.

        \paragraph{\textbf{Nanofoam}}
        We choose \SI{460}{\kelvin} for equilibration of the nanofoam, since it is well above $A_f$.
        Slices through the nanofoam as-created and after equilibration at \SI{460}{\kelvin} are shown in \Cref{fig:foam_temperature_cycle} (a) and (b).
        During equilibration the nanofoam is stable but the solid fraction and the ligament size slightly increase by less than \num{0.02} and \SI{0.05}{\nano\meter}, respectively.
        Each nanofoam is initially created as austenite B2 phase, but during equilibration about \SI{10}{\percent} of this phase transforms into the martensite phase.
        This happens preferentially close to surfaces and at points with small cross-section where the influence of the surface is pronounced.

        This is in line with the observations made in the nanowires, \Cref{fig:nanowire_size_effect} (b), where a decrease in the local length scale leads to an increase in $A_f$.
        Thus, the \SI{460}{\kelvin} are not sufficient to prevent the very small foam features from turning into martensite.
        Put in other words, surface effects favor martensite formation.

        Observations of the influence of free surfaces on structure in nanofoams are known from literature on NPG.
        Initial surface stresses pre-stress the material and lead to minor amounts of deformation and defects \cite{Ngo2015}.
        While this plastic deformation in NPG is accounted for by dislocations, in NiTi the austenite-martensite phase transformation is active to accommodate mechanical loads.
        Dislocations or stacking faults could not be found in the nanofoam, while in the nanowires at least martensite twinning was a prominent feature, see \Cref{fig:domain_boundaries_small_wire_b2_T_cycle}.
        Note, that dislocations in NPG are a defect characteristic of permanent plastic deformation representing irrecoverable damage.
        In contrast, the martensitic transformation in NiTi is reversible meaning that transformed regions are not irrecoverably damaged.
        In the next section, we use temperature to drive the reversible phase transformation in NiTi.

    \subsection{Temperature cycle}

        \begin{figure*}[htbp]  
            \centering
            \includegraphics{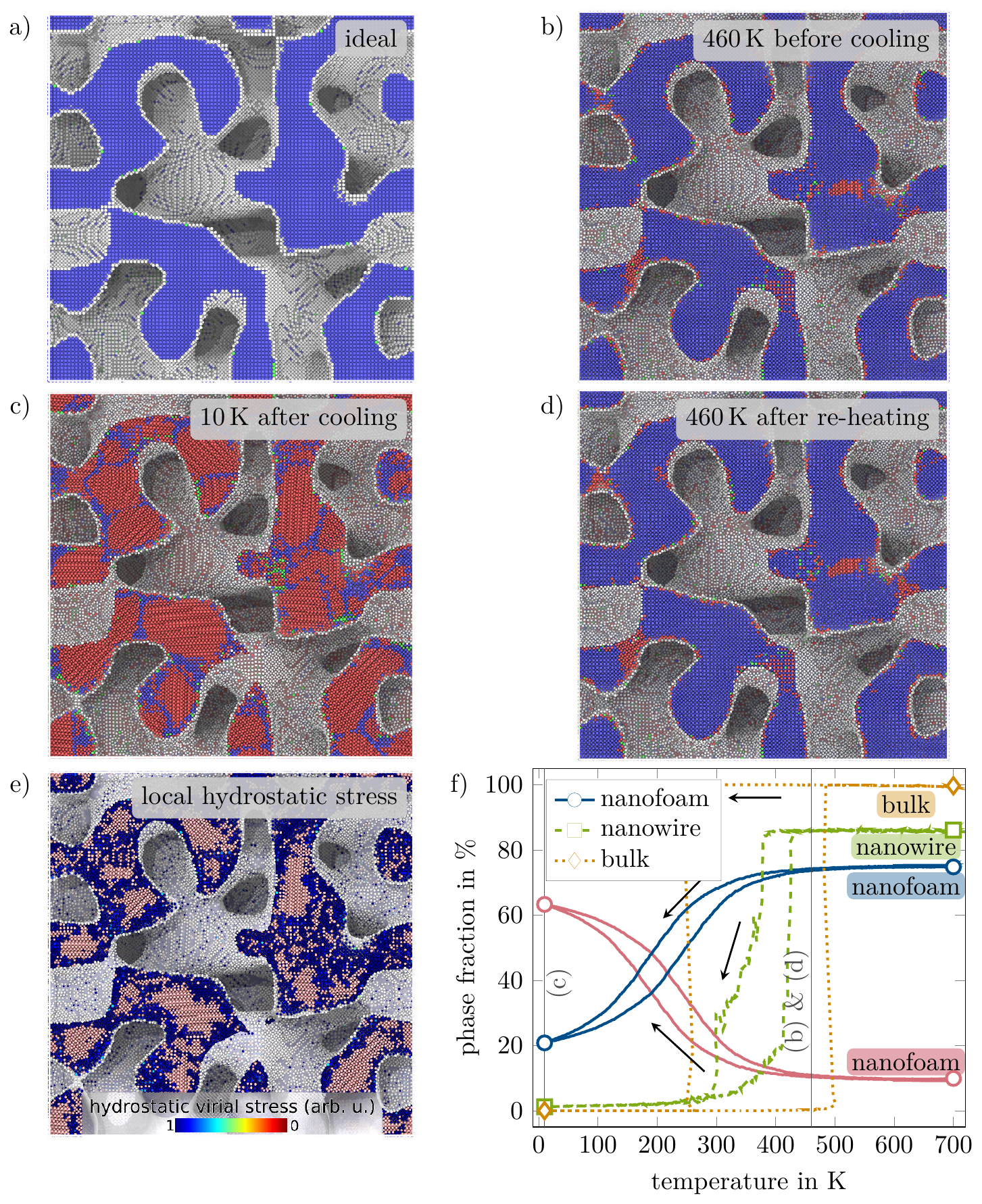}
            \caption{
                (a)-(d) Slices through a simulation snapshot from construction and throughout a temperature cycle. Blue = austenite, red = martensite, green = other phases, gray = surface atoms.
                (e) Situation as in (c) but with blue color coding for high absolute of hydrostatic per-atom stress. 
                (f) Fraction of crystallographic phases during temperature cycle of bulk, nanowire (\SI{8.4}{\nano\meter} diameter), and nanofoam with $\phi = 0.5$. Arrows indicated the cooling branch of the thermal cycle. Austenite and martensite phase fractions can both be used to track the state of the phase transformation.
                }
        \label{fig:foam_temperature_cycle}
        \end{figure*}

        We, first, compare bulk, nanowire, and nanofoam in the light of the so-called \textit{isobaric shape-memory effect}, i.e., the phase transformation in a temperature cycle is investigated.
        The stress-driven phase transformation, i.e., the \textit{isothermal superelastic effect} is left for the next section.
        As a representative example we shall focus on a nanofoam with  $\phi = 0.5$ and $L = \SI{4.1}{\nano\meter}$, see \Cref{fig:foam_temperature_cycle}.

        \paragraph{\textbf{Sample average}}
        The temperature driven phase transformation is fully reversible under stress-free conditions for all tested material configurations, which can be tracked by either the austenite or the martensite phase fraction.
        For the bulk material and large nanowires the transformation happens very suddenly upon reaching the critical transformation temperatures.
        Moreover, the surfaces reduce the spread between start and finish temperatures due to nucleation sites at the surface.
        With nanowire size the hysteresis becomes smaller, and the phase transition smears out.
        For the nanofoam, the phase hysteresis finally becomes a smooth curve without sudden insets of phase transformation.
        The jumps in the phase fraction curves characteristic for the cooperated phase transformation have disappeared completely.
        Also, its phase transformation does not proceed to full completion.
        Additionally, the start of the hysteresis curve for the nanofoam is shifted to lower temperatures compared to nanowires with similarly sized features (\Cref{fig:nanowire_size_effect,fig:foam_temperature_cycle}).
        Compared to bulk NiTi, the nanofoam offers smaller hysteresis and a more continuous phase transformation behavior at the cost of reducing the accessible amount of material subjected to phase transformation.

        \paragraph{\textbf{Local phase transformation}}
        The martensite that is retained at high temperatures primarily appears in regions where the local length scale is reduced, i.e., in ligaments with very small diameter or necks, see \Cref{fig:foam_temperature_cycle} (c).
        This is caused by the influence of surface tension that drives the phase transformation to lower temperatures, see \Cref{fig:nanowire_size_effect} (b).
        On the other end of the spectrum the austenite retained at low temperatures is primarily present in between domains of different martensite orientation where it is mechanically clamped, see \Cref{fig:foam_temperature_cycle} (e).

        During cooling the martensite can nucleate at various places in the structure simultaneously because the many differently oriented surfaces offer efficient nucleation sites.
        Note, that the initial structure is essentially a porous single crystal.
        Yet, due to the high symmetry of the austenite parent phase the martensite can form in different orientation variants.
        As the stimulus for phase transformation (i.e., temperature) is isotropic, none of these variants is preferred, and they all form spontaneously.
        As these martensite nuclei grow they approach other nuclei from neighboring regions.
        However, different martensite variants, i.e., martensite with different orientation of its local crystallographic axes do not necessarily have matching orientations creating grain boundaries/domain boundaries.
        Between two martensite variants, austenitic structure can be retained.

        In summary, the microstructure in temperature driven phase transformation features multiple martensite variants that form a domain-like pattern.
        Since various martensite variants nucleate simultaneously, regions in between the martensite domains remain as untransformed austenite.

    \subsection{Uniaxial compression}

        After having studied the isobaric shape-memory effect we turn to studying the behavior under applied strain at constant temperature, i.e., the isothermal superelastic effect.

        \paragraph{\textbf{Compression simulation}}
        Nanofoams of varying solid phase fraction $\phi$ are compressed along z-direction to strains of \SI{50}{\percent} and higher.
        The lateral directions (x- and y-direction) are kept stress-free.
        To obtain superelastic behavior the temperature is held above $A_f$ at \SI{460}{\kelvin} \cite{Shaw1995}. 
        Below we use the common neighbor analysis algorithm (CNA) instead of the PTM algorithm to identify phases since it is more reliable and efficient at strains exceeding \SI{10}{\percent}.

        \begin{figure}[htbp] 
            \centering
            \includegraphics{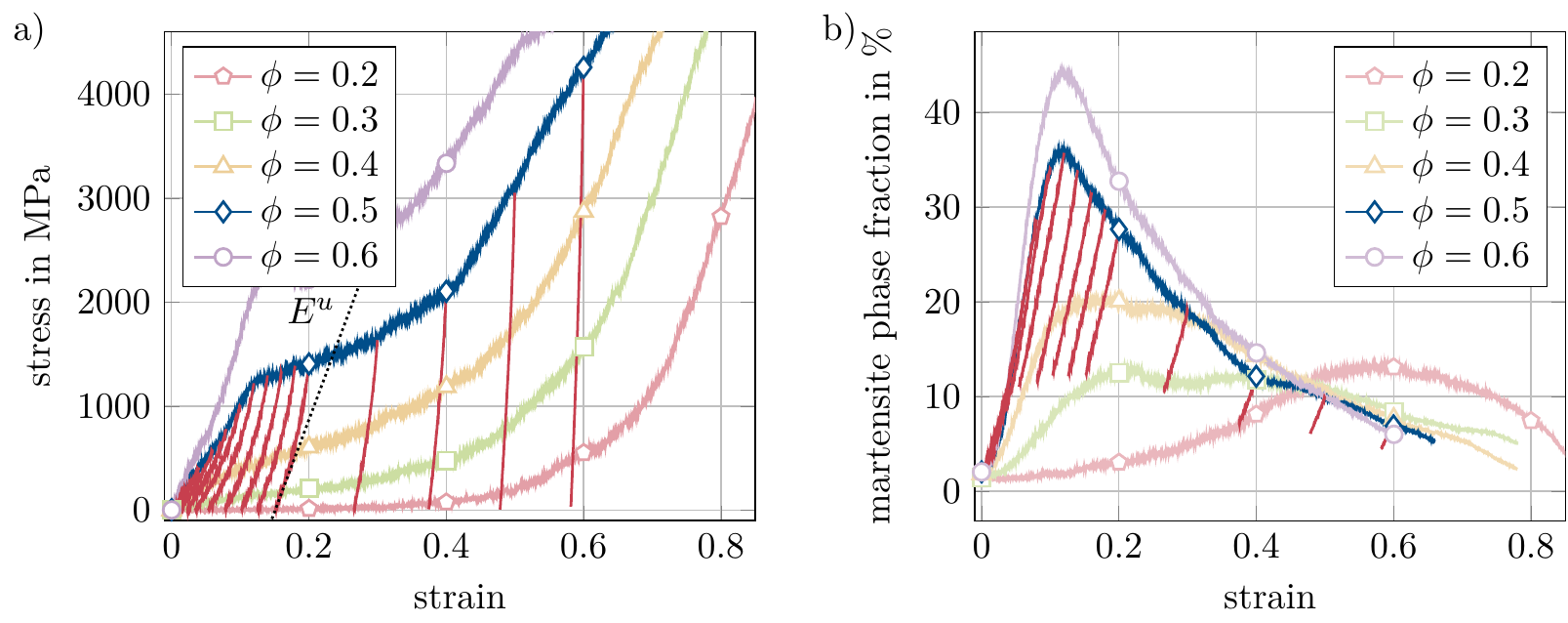}
            \caption{
                Stress-strain curves (a) and martensite fractions (b) of foams with different solid phase fractions $\phi = 0.2$ to $0.6$ during compression.
                For $\phi = 0.5$ the unloading curves from various points in strain are shown as an example.
                Up to \SI{10}{\percent} strain stress behaves nearly linear before compliance increases significantly.
                The tangent modulus in the unloaded state \(E^u\) is shown as an example for one unloading curve.
            }
            \label{fig:nanofoam_var_phi_compression}
        \end{figure}

\begin{figure}[htbp] 
            \centering
            \includegraphics{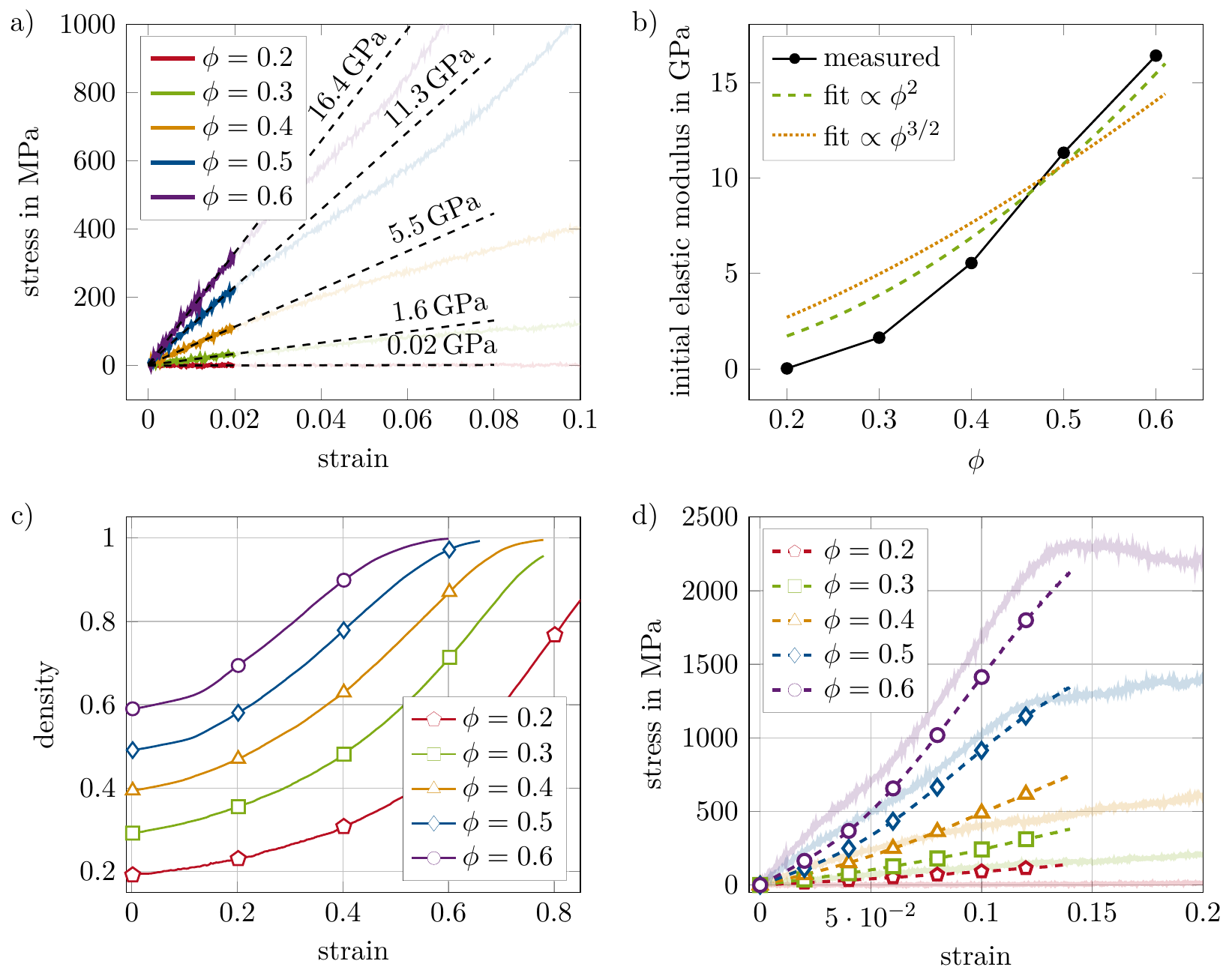}
            \caption{
                (a) Excerpts from the stress-strain curves indicating the linear region used for fitting of the initial elastic modulus.
                The respective modulus for nanofoams with solid phase fractions $\phi = 0.2$ to $0.6$ are indicated and plotted in (b).
                (b) Attempted fits the elastic moduli with different fit functions.
                (c) Density over compressive strain for foams with different solid phase fractions.
                (d) Prediction (dashed) of the measured initial stress-strain curve (shaded) with the equations \eqref{eq:eff_linear_elasticity} to \eqref{eq:eff_strain}.
                }
            \label{fig:nanofoam_stress_prediction}
        \end{figure}

        \paragraph{\textbf{Pseudo-elastic stress-strain curves}}
        In \Cref{fig:nanofoam_var_phi_compression} (a) the loading stress strain curves and select unloading curves for \(\phi=0.5\) are shown.
        It appears that the stress increases linearly during the initial \SI{2}{\percent} strain.
        Beyond this point the curve becomes slightly curved up to about \SI{11}{\percent} compressive strain.
        Therefore, we fit a linear regression to the initial \SI{2}{\percent} strain and plot the moduli extracted from the slopes, see \Cref{fig:nanofoam_stress_prediction} (a) \& (b).
        Neither fitting the modulus with a \(\phi ^2\) trend as suggested by the Gibson-Ashby scaling laws for the Young's modulus \cite{Gibson1982}, nor with a \(\phi^{3/2}\) trend as suggested for the fracture stress of NPG in Ref. \cite{Jin2018}, gives satisfactory agreement.

        \paragraph{\textbf{Elastic modulus in the light of phase transformation}}
        As alternative, we try a more elaborate approach to explain the stress-strain curves below \SI{11}{\percent} strain, which we shall call regime A.
        Phase transformation commences immediately as stress is applied, see \Cref{fig:nanofoam_var_phi_compression} (b).
        The quasi-linear region at strains below \SI{11}{\percent} is, thus, not the result of elastic deformation, but appears despite the superelastic martensite formation.
        Consequently, it is important to take into account the correct phase fractions when describing the stress \(\sigma\) as a function of effective strain \(\epsilon^{\prime}\).
        Assume that the rules of linear elasticity apply,
        \begin{align}
            \sigma = Y^{\rm eff} \epsilon^{\prime} \text{,} \label{eq:eff_linear_elasticity}
        \end{align}
        and that the elastic modulus \(Y^{\rm eff}\) follows a Gibson-Ashby behavior:
        \begin{align}
            Y^{\text{eff}} = Y^{\text{bulk}} \phi^2 \text{.} \label{eq:Gibson_Ashby}
        \end{align}
        However, both \(\phi\) and \(Y^{\text{bulk}} \) may vary with strain.
        We can easily measure the true solid volume fraction \(\phi\); see \Cref{fig:nanofoam_stress_prediction} (c).
        For \(Y^{\text{bulk}} \) we assume a simple linear combination of the elastic moduli of the austenite and the martensite phases,
        \begin{align}
            Y^{\text{bulk}} = \phi_M Y^{\text{M}} + \phi_A Y^{\text{A}} \text{,} \label{eq:linear_combination_modulus}
        \end{align}
        using the phase fractions of austenite \(\phi_A\) and martensite \(\phi_M\), and the elastic parameters \(Y^{\text{M}}=\SI{90}{\giga\pascal}\) and \(Y^{\text{A}}=\SI{31}{\giga\pascal}\) of the interatomic potential \cite{Ko2015}.
        In addition, the strain upon phase transformation from austenite to martensite \( \epsilon_{\rm phase-change} \) is taken into account, 
        \begin{align}
            \epsilon_{\rm phase-change} = ( \phi_M - \phi_M^{\rm initial} ) * \epsilon_{A \rightarrow M} \text{,} \label{eq:phase_change_strain}
        \end{align}
        and used to obtain \(\epsilon^{\prime} \):
        \begin{align}
            \epsilon^{\prime} = \epsilon - \epsilon_{\rm phase-change} \text{.} \label{eq:eff_strain}
        \end{align}
        For \(\epsilon_{A \rightarrow M}\) a value of \num{0.04} is inserted, computed for the given interatomic potential \cite{Ko2015}.
        
        Using these formulas and the obtained data we predict \(\sigma\) and compare it with the measured values during the simulation in \Cref{fig:nanofoam_stress_prediction} (d).
        The agreement between our model for \(\sigma\) and the simulation is reasonable.
        Our model is overestimating the stiffness for nanofoams of low solid phase fraction but is underestimating it at higher phase fractions and low strains.

        On the one hand, the discrepancy for low \(\phi\) is due to dangling ligaments/appendices, see Refs. \cite{Jin2018,Klomp2021} and references therein.
        The low pseudo-elastic modulus of the low \(\phi\) foams is in line with these samples showing a lower fraction of martensite formed during the compression.
        Less material loaded, means there is less material that is driven into the martensite phase.
        We, therefore, expect that the phase transformation only occurs locally and not within the whole sample as in the case of the temperature cycle, see \Cref{fig:foam_temperature_cycle}.
        On the other hand, the fact that deviations grow as strain increases is attributed to the onset of deformation mechanisms that go beyond elasticicty and austenite to martensite phase transformation.
        We will discuss these mechanisms below.

        \paragraph{\textbf{Deformation at intermediate strain}}
        The shape of the curves in \Cref{fig:nanofoam_var_phi_compression} (b) indicates that there is a point of maximum phase transformation that depends on \(\phi\).
        If strained further the martensite phase fraction decreases again.
        At around \SI{10}{\percent} strain, alternative mechanism of deformation set in.
        Therefore, we define regime B in the range from \SI{5}{\percent} to \SI{14}{\percent} strain for the \(\phi = 0.5\) foam, which overlaps with regime A.

\begin{figure}[htbp] 
            \centering
            \includegraphics{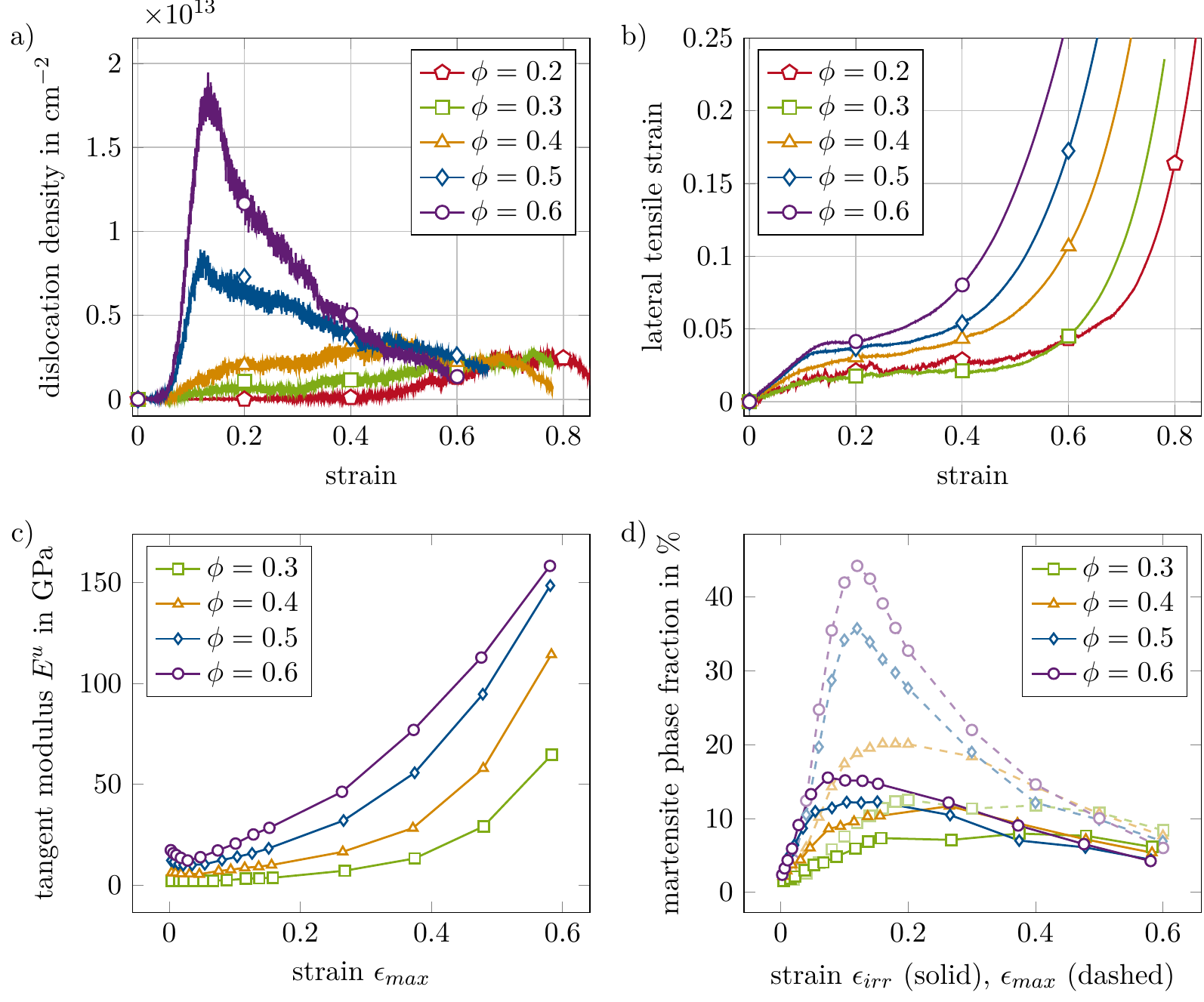}
            \caption{
                (a) Total dislocation line density over strain obtained with Ovito's DXA.
                Note that also twin boundaries are detected as groups of dislocations.
                (b) Lateral (tensile) strain over applied (compressive) strain for foams with different solid phase fractions.
                The values have been symmetrized, i.e., \(\epsilon_{\rm lateral} = \frac{1}{2} \left(\epsilon_{xx} + \epsilon_{yy}\right)\).
                (c) Tangent modulus of the zero stress state after complete unloading from different applied maximum strain $\epsilon_{max}$.
                The \(\phi = 0.2\) foam has been excluded because the stress levels are too low compared to thermal fluctuations to obtain reliable tangents from \Cref{fig:nanofoam_var_phi_compression} (a).
                (d) The fraction of martensite phase at the point of maximum applied strain is plotted over the maximum applied strain (dashed); and the fraction of martensite phase after unloading is plotted over the irreversible strain (solid).
                Data for different phase fractions is obtained by CNA and can be directly compared to (c).
            }
            \label{fig:nanofoam_damage_and_high_strain}
        \end{figure}

        The decreasing amount of material identified as martensite (\Cref{fig:nanofoam_var_phi_compression} (b)) is due to less phase recognized by the CNA algorithm as crystalline.
        It may either be a very distorted martensite phase or the sample has been amorphized.
        We recognize, that this may also be an artifact of the high strain rates in the simulation.
        As an additional indication that the analytical description breaks down here, we track the dislocation density in the sample in \Cref{fig:nanofoam_damage_and_high_strain} (a).
        It shows that dislocation activity is prominent in the very region where the above analytic scaling breaks down.
        From the plot of density over strain, \Cref{fig:nanofoam_stress_prediction} (c), and lateral strain over applied strain, \Cref{fig:nanofoam_damage_and_high_strain} (b), we see further compression but vanishing lateral expansion.

        These observations indicate that the deformation now has become irreversible and neither strain nor phase fraction can recover, see \Cref{fig:nanofoam_var_phi_compression}.
        A summary of the irreversibility of the phase transformation and strain is displayed in \Cref{fig:nanofoam_damage_and_high_strain} (d).

\begin{figure*}[htbp] 
            \centering
            \includegraphics{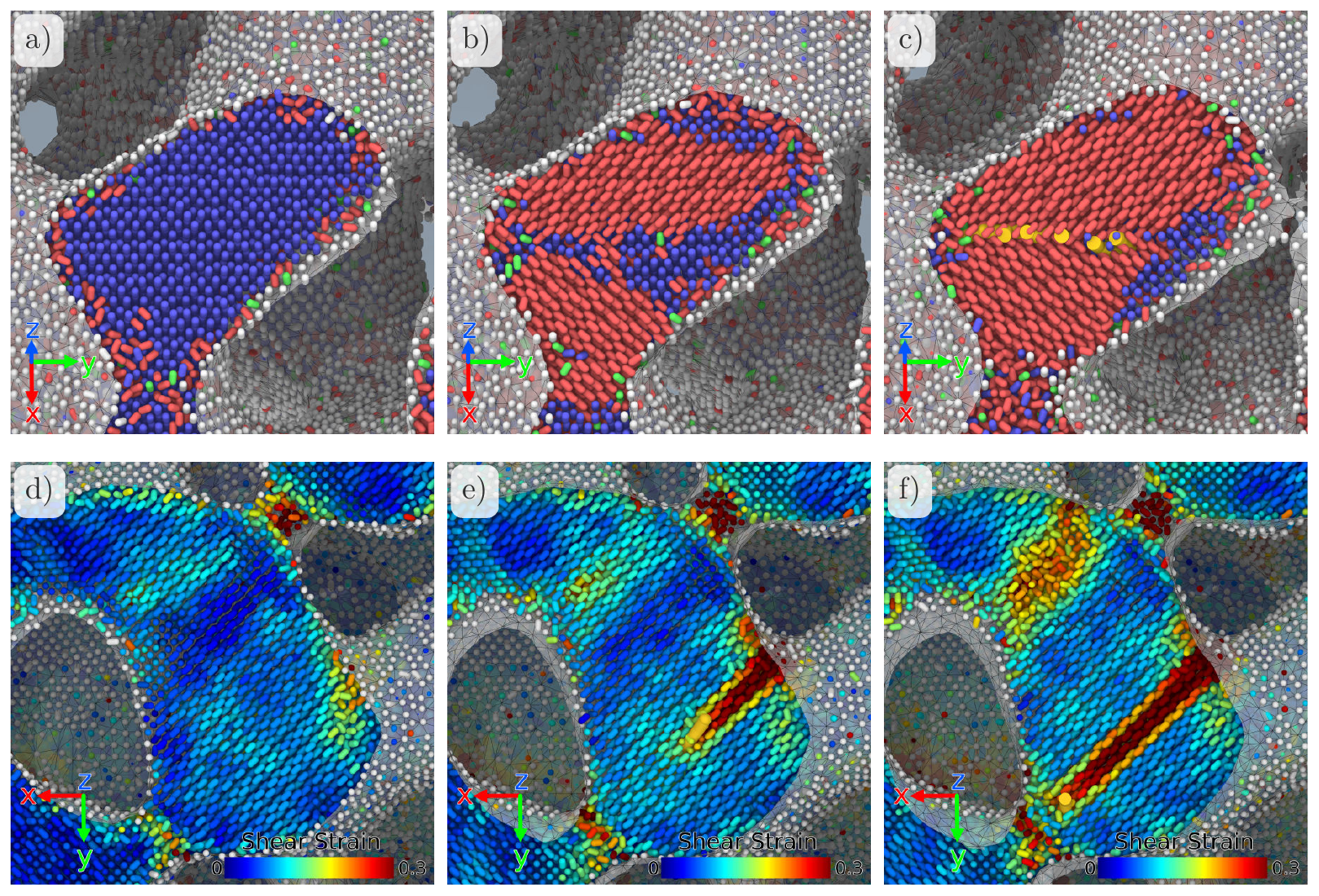}
            \caption{
                Top row (a) - (c): formation of two martensite variants with a twin boundary.
                Color coding by crystal phase, dislocations are shown in orange.
                (a) \(\epsilon = \SI{0}{\percent} \),
                (b) \(\epsilon = \SI{8}{\percent} \),
                (c) \(\epsilon = \SI{17}{\percent} \). 
                Bottom row (d) - (f): accumulation of shear strain by dislocations and amorphization
                Color coding by atomic shear strain, dislocations are shown in orange, in the top part of the ligament we see an amorphized region.
                (d) \(\epsilon = \SI{10}{\percent} \), 
                (e) \(\epsilon = \SI{14}{\percent} \), 
                (f) \(\epsilon = \SI{15}{\percent} \).
                }
            \label{fig:nanofoam_damage_microscopic}
        \end{figure*}

        The processes that are at interplay in this region are a decreasing amount of phase transformation, formation of twinned structures, dislocation glide and amorphization of crystal structure that all occur while densification speeds up.
        Examples for the relevant microscopic processes are shown in \Cref{fig:nanofoam_damage_microscopic}.

        In \Cref{fig:nanofoam_damage_microscopic} (a) - (c) snapshots from the \(\phi = 0.5\) foam show how two martensite variants form in different locations of an initially austenitic ligament.
        The martensite grows until the regions of the two variants come very close.
        In between there is still austenite phase retained, which is mechanically clamped by the two martensite regions.
        Only at significantly higher strain a kind of twin boundary is formed.
        The region that forms the twin boundary resembles a network of dislocations (orange structure).
        Therefore, these dislocations are an artifact that obscures the interpretation of the total dislocation density making it difficult to interpret \Cref{fig:nanofoam_damage_and_high_strain} (a) in the light of dislocation-based plasticity.

        In the second series of snapshots in \Cref{fig:nanofoam_damage_microscopic} (d) - (f) the atoms are color coded by the local shear strain they have experienced in the course of the compression simulation revealing two other effect that occur upon deformation.
        The lower part clearly shows how a quantum of plastic deformation in the crystal is carried by a dislocation (orange).
        The trace of the dislocation is clearly visible by the high atomic shear strain.
        However, in the top part of the image there is a region where the shear strain increases gradually, apparent from the gradual change in color from blue to orange.
        In this region, the material's crystal structure is no longer recognized by the CNA algorithm.
        Thus, we suspect that locally the material has become amorphous.
        This represents another form of accommodation of plastic strain in the material.

        Note that the trends are identical for the lower density foams, but that all the regions are much more smeared and set in later.

        \paragraph{\textbf{Deformation at high strain}}
        The regimes of high strain are most easily identified in the plot of lateral strain, see \Cref{fig:nanofoam_damage_and_high_strain} (b).
        We define region C as the plateau appearing in lateral strain.
        This plateau is shorter the higher the initial \(\phi\).
        In this region we find a high density of dislocations, see \Cref{fig:nanofoam_damage_and_high_strain} (a), which suggests that dislocation and twinning activity is high.
        As soon as a true density of around \SI{60}{\percent} to \SI{70}{\percent} is reached, see \Cref{fig:nanofoam_stress_prediction} (c), the plateau ends and we are in regime D.
        Here stress (\Cref{fig:nanofoam_var_phi_compression} (a)) and lateral tensile strain (\Cref{fig:nanofoam_damage_and_high_strain} (b)) show an accelerated increase.
        Pores have mostly closed, and the material is pressed out to the sides.

        \paragraph{\textbf{Summary}}
        Finally, let us summarize the findings on the elasto-plastic behavior of the NiTi-nanofoams of the example of the \(\phi = 0.5\) foam in regions A to D
        \begin{enumerate}[label=\Alph*.]
            \item Up to approx.~\SI{5}{\percent} strain deformation is dominated by phase transformation.
            \item Up to approx.~\SI{14}{\percent} there is an interplay of phase transformation, twinning, dislocation glide, and amorphization while the densification slightly speeds up.
            \item We seem to have a speeding up of the densification and there is more amorphization. Also, high dislocation glide activity and twinning occurs.
            \item Due to massive compaction at high strain, the sample looses its crystal structure as pores are nearly closed.
        \end{enumerate}
        While regions A and B are inherent to the superelastic properties of NiTi, regions C and D are analog to the behavior of NPG.

    \subsection{Unloading from compression}

        After compression, the foams are unloaded to zero stress.
        Intermediate snapshots of the loading simulation are taken and unloaded separately, such that different samples have experienced different maximum strains $\epsilon_{max}$.

        The unloading stress-strain curves and phase fraction-strain curves are exemplarily shown for the $\phi = 0.5$ nanofoam in \Cref{fig:nanofoam_var_phi_compression}.
        From these stress-strain curves the tangent modulus $E^u$ can be extracted.
        It serves as a proxy for the elastic modulus in the pseudo-elastic/superelastic setting of the unloaded state.
        In addition, we define the strain and martensite phase fraction retained after unloading as the irreversible strain \(\epsilon_{irr}\) and the irrecoverable martensite fraction, respectively.

        \paragraph{\textbf{Reversibility}}
        Again focussing on the \(\phi=0.5\) nanofoam, the amount of irrecoverable martensite phase and irreversible strain are proportional to each other at low strains, see \Cref{fig:nanofoam_damage_and_high_strain}~(d).
        The increase in irrecoverable martensite is quite high (steep slope) but also the reversibility part increases steeply (dashed curve).
        Incomplete reversal of the phase transformation is the prime reason for incomplete shape recovery in the low strain regime A, analog to macroscopic NiTi foams.
        Enhancement of shape recovery is expected at higher temperatures which bring the thermodynamic behavior more into the superelastic regime.

        As strain is increased, especially beyond the \SI{10}{\percent} limit, the irreversible strain increases over proportion (the curve flattens) with respect to the maximum as well as the irrecoverable martensite fraction.
        This is a direct consequence of the above discussed mechanisms of deformation, which change at the transition from deformation regime A to B.
        When less strain is carried by a reversible phase transformation also the reversibility of the nanofoam degrades.
        Irreversible processes such as coalescence of structural elements and failure of foam ligaments increase in importance beyond \SI{10}{\percent} strain.            
        Thus, if good reversibility shall be achieved, loading should be limited to \SI{10}{\percent} strain, such that the irreversible strain $\epsilon_{irr}$ stays at around \SI{4}{\percent} or less in the large $\phi$ samples.

        \paragraph{\textbf{Modulus $E^u$ and phase composition}}
        In \Cref{fig:nanofoam_damage_and_high_strain} (c) we see that the tangent modulus $E^u$ first shows a slight decrease around \SI{3}{\percent} to \SI{5}{\percent} and afterwards increases continuously.
        The continuous increase at high strains is in line with the above observations in phase transformations and damage.

        In the minimum at low strains there exists a combination of austenite and martensite phase in the foam, see \Cref{fig:nanofoam_damage_and_high_strain} (d).
        Therefore, there are always enough nucleation centers for a martensitic as well as an austenitic phase transformation.
        Additionally, there is always some austenite as well as some martensite that is in a pre-stressed state such that an infinitesimal change in strain leads to some phase being transformed.
        Thus, the apparent modulus is decreased since even the slightest change in load leads to apparent compliance.

    \subsection{Cyclic compression and decompression}

        From the single loading-unloading cycle in the previous section it is evident that, even below the \SI{10}{\percent} strain limit (regime A) shape recovery is not ideal, see \Cref{fig:nanofoam_damage_and_high_strain} (d).
        Irreversible deformation quantified by the irreversible strain and related to irrecoverable martensite is remnant after unloading.
        This suggests that, the phase transformation is less reversible compared to nanowire or bulk samples.
        But does the damage accumulate when multiple loading-unloading cycles are applied?
        The cyclic loading scenario is especially important to cyclic applications such as actuation or elastocaloric applications. 

        Experimentally, cyclic compressive loading in macroscopic samples was studied with varying number of load cycles \cite{Yuan2004, Yuan2005, Sakurai2006, Zhang2008, Guo2009, Aydogmus2012, Xu2015, Zhang2015}.
        After an initial training of 1 to 5 cycles near-perfect superelasticity could be observed for low applied strains.
        For even higher cycle numbers (reports vary between \num{20} cycles to \num{230000} cycles) plastic deformation slowly accumulated, microcracks evolved and samples finally fractured \cite{Aydogmus2012,Bram2011}.

        \paragraph{\textbf{Simulation of cyclic load}}
        To this end, the $\phi = 0.5$ foam is subjected to cyclic loading and unloading.
        We limit ourselves to the first few loading-unloading cycles when damage accumulation is strongest.
        The loading condition can be described as a compressive ripple strain with $R=-\infty$ and constant maximum strain.
        In total five different maximum strain levels are tested, but for clarity only a representative selection is shown.

        \begin{figure}[htbp] 
            \centering
            \includegraphics{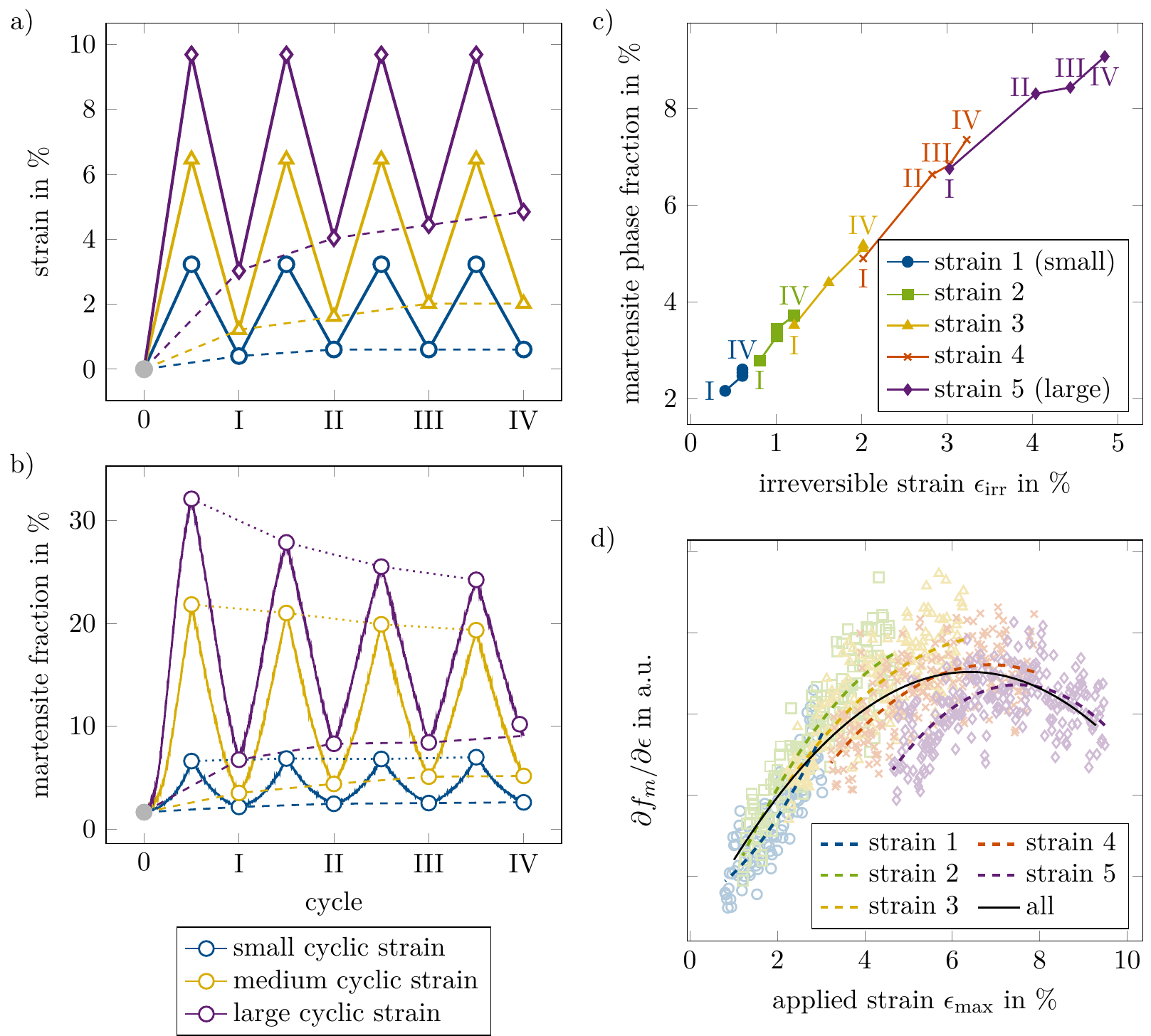}
            \caption{
                Identical $\phi=0.5$ nanofoam samples have been compressed to various maximum strains and then unloaded to zero stress at \SI{460}{\kelvin} for 4 consecutive cycles 0 to IV.
                (a) Accumulation of irreversible strain for three different levels of maximum applied strain.
                (b) Accumulation of irrecoverable martensite (dashed lines); and simultaneous decrease of maximum martensite fraction achievable (dotted lines).
                (c) Remnant martensite and irreversible strain for different maximum strains.
                Each connected set of dots represents the complete 4 cycles of one sample from (b).
                The first load cycle in each set is represented by the lowest and leftmost point.
                Damage to the nanofoams is more pronounced for increased maximum strain.
                (d) Change in martensite phase fraction as a function of applied strain.
                Plot of the martensite phase fraction derived with respect to strain the fourth compression.
                Around \SI{6}{\percent} the largest amount of phase transformation happens in a given interval of strain.
                Fit curves are regressions with third order polynomials.
            }
            \label{fig:cyclic_loading}
        \end{figure}

        \paragraph{\textbf{Irreversible strain}}
        The maximum strains chosen for the cyclic loading are all below the \SI{10}{\percent} strain limit identified above.
        Nevertheless, during removal of stress the nanofoams do not completely reverse their deformation, i.e. there is irreversible strain, see \Cref{fig:cyclic_loading} (a).
        The irreversible strain $\epsilon_{\text{irr}}$ is traced by the dashed lines.
        The higher the loading strain the worse is the strain recovery.
        Moreover, it appears that the major part of $\epsilon_{\text{irr}}$ is generated in the very first cycle and damage accumulation slows down significantly after that.
        Yet, a steady state is not achieved after four cycles.

        \paragraph{\textbf{Irrecoverable martensite}}
        The fraction of martensite phase during the cycles behaves in an analog fashion.
        Data from \Cref{fig:cyclic_loading} (c), therefore, suggests that irreversible strain and the amount of martensite present after unloading, i.e., remnant martensite, accumulate simultaneously.
        To clarify this proportionality, remnant martensite and irreversible strain are put into the same plot in \Cref{fig:cyclic_loading} (b).
        First, it shows excellent correlation between martensite phase fraction and irreversible strain.
        Therefore, as irrecoverable martensite accumulates, so does the strain that remains after unloading (irreversible strain).
        It also suggests that a more stable martensite phase, e.g. at higher temperatures, could lead to better shape recovery.
        The irreversible strain can, thus, be described in terms of phase composition.
        In contrast to e.g. NPG this strain is only due to phase transformation induced superelasticity.

        Second, the higher the maximum loading strain, the higher and faster the damage accumulation.
        This is in line with experimental and our theoretical observations \cite{Yuan2004,Yuan2005,Sakurai2006,Zhang2008,Guo2009,Bram2011,Aydogmus2012,Xu2015,Zhang2015}.

        \paragraph{\textbf{Comparison to NPG}}
        In nanofoams made of simple ductile metals such as NPG plastic deformation accumulates quickly and is permanent.
        By contrast, NiTi nanofoams can recover significant amounts of strain after loading.
        Although, recovery is not ideal, superelasticity allows for a much greater range of reversibly applied load.
        Ultimately, it could, therefore, have a much wider range of application in e.g. actuation or show a better tolerance against mechanical damage.
        Additionally, phase transformations allow an added degree of freedom when it comes to modified surface properties of the nanofoam that could be applied in catalysis scenarios.

    \subsection{Rate of phase transformation}

        The limit for irreversible changes in the NiTi foam to occur was identified to be about \SI{10}{\percent} strain for the $\phi = 0.5$ nanofoam at \SI{460}{\kelvin} above.
        Below this limit we now seek to find an optimum operation regime.
        The idea is to keep the applied strain to a minimum while maximizing the amount of (possibly reversible) phase transformation.
        We, thus, need a measure for the effectiveness of applied strain with regard to transformed phase.
        Therefore, the derivative of martensite phase fraction with respect to strain, ${\partial f_{m}}/{\partial \epsilon}$, which is essentially the slope of the plot in \Cref{fig:nanofoam_var_phi_compression} (b), is calculated.
        Instead of taking the first loading cycle, we take the data from the compression part of the fourth and last cycle of the cyclic loading.
        It is expected to be closest to a steady state configuration.

        Apparently there is a distinct maximum in the derivative of phase fraction with respect to strain at around \SI{6}{\percent} strain, see \Cref{fig:cyclic_loading}~(d).
        Here the martensite phase fraction changes fastest for a given change in applied strain.
        Below this optimum, elastic deformation has a strong contribution.
        Above, the martensite introduced in the structure does not revert to austenite upon unloading or becomes permanently deformed.
        Consequently, we propose to use ${\partial f_{m}}/{\partial \epsilon}$ as a figure of merit when it comes to optimal exploitation of the superelastic properties of NiTi.

\section{Discussion}

    The mechanical behavior of NiTi nanofoams has been addressed using molecular dynamics simulations.
    In a systematic approach the reactions of the nanofoam to thermal and mechanical loads have been studied and contrasted with bulk and nanowire behavior.
    The chosen approach of using molecular dynamics simulations instead of micromechanical models to study NiTi foams has proven suitable and reliable at providing information on the microscopic interplay of the austenite and martensite phase.

    In the light of the behavior of other nanoporous metals, especially NPG, some features are clearly changed for the NiTi material system.
    This is congruent with our hypothesis that the possible superelastic/shape-memory effect of the base alloy alters the mechanical properties of the nanofoam.
    The high strain-hardening coefficient and zero plastic Poisson ratio could not be confirmed with NiTi.
    Instead of extensive formation of dislocations in early yielding phases, NiTi accommodates strain initially by martensitic phase transformation.
    This capability is, nevertheless, limited and saturates at strains exceeding \SI{10}{\percent}.

    Comparing to macroscopic NiTi foams we could confirm that beyond a certain limit of strain superelasticity fails.
    In simulation, the limit for good reversibility was identified to be around \SI{6}{\percent} strain and a strong increase of permanent irreversible deformation sets in at \SI{10}{\percent}.
    Yet, a steady state could not be attained after only four loading cycles.
    Neither an overstabilization of austenite nor the complete disappearance of the superelastic effect at very small sizes could be confirmed.
    Nevertheless, the decrease of hysteresis with decreasing feature size and also with structural randomness in the foam structures was apparent.

    In summary, nanoporous NiTi foams exhibit intriguing thermomechanical properties that extend the range of the currently known nanoporous metals.
    Superelasticity may, thus, improve functionalities that have been conceived for NPG such as actuation or catalysis as well as applications such as damping MEMS devices that employ microscopic NiTi parts.
    In the future an experimental realization of nanoporous NiTi would, therefore, be of great interest.

\section*{Acknowledgments}
    Financial support to Arne J. Klomp has been granted by the Deutsche Forschungsgemeinschaft via the SPP 1599.
    Calculations for this research were conducted on the Lichtenberg high performance computer of the TU Darmstadt.

\section*{Author Contributions}
        
    The work was conceived by Karsten Albe.
    Simulations and analyses were conducted by Arne J. Klomp.

\section*{Competing Interests statement}

    The authors declare no competing interests

    \bibliography{bib/complete_zotero_literature}

\end{document}